\documentclass[12pt,preprint]{aastex}
\usepackage{natbib}
\usepackage{amsmath}
\usepackage{graphicx}

\shorttitle{The New sdBV Star JL 166}
\shortauthors{Barlow, Dunlap, Clemens}

\begin{document}

\title{Detection of Photometric Variations in the sdBV Star JL 166\footnote{Based on observations at the SOAR Telescope, a collaboration between CPNq-Brazil, NOAO, UNC, and MSU.}}

\author{B.N. Barlow\footnote{Department of Physics and Astronomy, University of North Carolina, Chapel Hill, NC 27599-3255, USA; bbarlow@physics.unc.edu, bhdunlap@physics.unc.edu, clemens@physics.unc.edu} , B.H. Dunlap\footnotemark[2] , A.E. Lynas-Gray\footnote{Department of Physics, University of Oxford, Keble Road, Oxford OX1 3RH, England; aelg@astro.ox.ac.uk} , \& J.C. Clemens\footnotemark[2]}

\begin{abstract}
We report the discovery of oscillations in the hot subdwarf B star JL 166 from time-series photometry using the Goodman Spectrograph on the 4.1-m Southern Astrophysical Research Telescope.  Previous spectroscopic and photometric observations place the star near the hot end of the empirical sdB instability strip and imply the presence of a cool companion.  Amplitude spectra of the stellar light curve reveal at least 10 independent pulsation modes with periods ranging from 97 to 178 s and amplitudes from 0.9 to 4 mma.  We adopt atmospheric parameters of T$_{eff}$ = 34350 K and log \textit{g} = 5.75 from a model atmosphere analysis of our time-averaged, medium-resolution spectrum.
\end{abstract}
\keywords{stars: subdwarfs -- stars: oscillations -- stars: individual: JL 166}

\section{Introduction}
Hot subdwarf B (sdB) stars are a class of objects identified with models of extended horizontal branch stars and have temperatures ranging from 22000 to 40000 K and log \textit{g} values from 5.0 to 6.0 dex. They dominate surveys of faint blue objects at high galactic latitudes and are often cited as the main source of the UV excess observed in giant elliptical galaxies.  Although their origins and evolutionary tracks are still debated, the sdB stars are believed to be evolved, lower-mass (\~{}0.5$M_{\sun}$) stars with a He-burning core surrounded by a thin H layer.  Models of sdB stars show they will evolve directly to the white dwarf cooling sequence after core He exhaustion.   Their optical spectra are dominated by Balmer lines and sometimes also display He lines.

\citet{cha96,cha97} predicted the existence of a class of pulsating sdB stars based on the presence of p-modes in their model stars at the right temperatures and surface gravities.  The oscillations they found were driven by a $\kappa$-mechanism associated with an opacity bump from the ionization of Fe.  Contemporaneous with this prediction, \citet{kil97} reported the first detection of p-mode oscillations in the sdB star EC 14026-2647, opening up the possibility of using asteroseismological methods to probe the interiors of these stars.  Since this discovery, more than 60 pulsating sdB stars (sdBVs) have been observed with either p- or g-mode oscillations.  The rapid p-mode pulsators\footnote{referred to as V361 Hya or EC 14026 stars} oscillate with periods between 80 and 600 s and amplitudes around 10 mma.  The g-mode pulsators\footnote{also known as PG1716, V1093 Her, or Betsy stars} have periods ranging from 1 to 2 hrs and amplitudes comparable to the p-mode pulsators.  In the log \textit{g}-T$_{eff}$ plane, the p-mode oscillators cluster together in an instability strip with T$_{eff}$ from 28000 to 35000 K and log \textit{g} between 5.2 and 6.1 dex.  The slow pulsators typically have cooler temperatures and lower surface gravities with T$_{eff}$ between 23000 and 30000 K and log \textit{g} near 5.4 dex.  In the region where the red edge of the p-mode instability strip overlaps the blue edge of the g-mode pulsator strip, three sdBVs have been discovered that show both p- and g-mode oscillations \citep{sch06,bar06,lut09}.

We report the discovery of p-mode oscillations in JL 166, a hot subdwarf B star that was first listed in the catalogue of \citet{jai69}.  In their survey, \citet{jai69} used the Schmidt telescope of the Uppsala Southern Station at Mount Stromlo to search for faint violet stars in Southern galactic latitudes.  296 stars were catalogued in the survey; their colors were assigned ``decidedly violet'' or ``possibly violet'' designations, while their brightnesses were estimated as ``bright,'' ``intermediate,'' or ``faint.''  JL 166 was classified as ``decidedly violet'' with ``intermediate'' brightness, and further observations with a 100 cm reflector at Siding Spring Observatory led to B and V measurements of 15.00 and 15.23, respectively \citep{jai69}.  More recent 2MASS observations \citep{skr06}\ found J, H, and K magnitudes that imply the presence of a cool companion.  The only published spectral data for JL 166 \citep{heb86}\ revealed strong Balmer lines, He I lines, and the 4686 \AA\ He II line, and resulted in an sdOB spectral classification.  \citet{heb86}\ reported temperature and gravity values of 35500 K and 5.8 dex, respectively, which place JL 166 near the blue edge of the empirical sdB instability strip (see Fig. 1 of \citealt{cha01}).

We first learned of JL 166 after searching the online Subdwarf Database \citep{ost06} for sdB stars in the instability strip.  Noticing JL 166 fell in the p-mode instability strip, we observed the star to look for variations in the stellar luminosity.  With the Goodman Spectrograph on the 4.1 m Southern Astrophysical Research (SOAR) telescope, we have detected at least 10 independent frequencies in the light curve with periods ranging from 97 to 178 s and amplitudes from 0.9 to 4 mma.

\citet{heb09} reviews what is currently known about sdB stars, a poorly understood late-stage of stellar evolution.  In particular, \citet{heb09} highlights the diversity and emphasizes (his \S 14) the need for more determinations of T$_{eff}$ and log \textit{g}.  Further studies, such as the one on JL 166 presented here, would in due course permit a better, informed study of sdB star evolution and their contribution to the ultraviolet upturn seen in the spectra of giant elliptical galaxies.

\section{Photometry}

We obtained time-series photometry for JL 166 with the Goodman Spectrograph on the 4.1-m SOAR telescope.  In imaging mode, the camera-collimator combination re-images the SOAR telescope focal plane with a focal reduction of three times, resulting in a plate scale of 0.15 arcsec pixel$^{-1}$.  The camera houses a 4k x 4k Fairchild back-illuminated CCD with electronics and dewar provided by Spectral Instruments, Inc.  Using optics of fused silica, NaCl, and CaF$_{2}$, the entire system is optimized for high throughput from 320 to 850 nm. See \citet{cle04} for further details on the spectrograph. 

We observed JL 166 on two engineering nights in 2008 September and October, obtaining uninterrupted runs of over 2 hours on each night.  Each run was obtained through a red-blocking S8612 filter, which has a bandpass of 300 to 700 nm.  To minimize the processing time between exposures yet keep the readnoise low, we read out a small subsection of the CCD (approximately 500 x 500 pixels binned 2x2) and used a readout speed of 100 kHz.  Each exposure had an integration time of 10 s, resulting in a cycle time of \~{}13 s.  Table 1 summarizes the details of our photometric observations.  

In order to obtain light curves from the raw images, we extracted our photometry using the external IRAF package CCD\_HSP developed by Antonio Kanaan, which employs the aperture photometry preferred by \citet{odo00}.  Aperture widths were chosen to maximize the signal-to-noise ratio in the light curves and were approximately 1.7 times the seeing width.  To subtract the sky from each stellar aperture, we used sky annuli with widths of 1.5 arcsec that started approximately 4.5 arcsec from the centers of the stars.  We divided our light curves by those of a constant comparison star to remove small-scale variations in the sky transparency.  The comparison star used is located approximately 02\arcmin\ 39\farcs4 southwest of JL 166 and was the only one present in our frames with an adequate signal-to-noise ratio.  Atmospheric extinction effects were corrected by fitting and normalizing the curves with parabolas.  We note that this normalization may remove real variations in the stellar brightness on the order of our run length.  Due to superb telescope guiding, we did not flat-field or bias-subtract any of the frames.  The reduced light curves for JL 166 are presented in Figure 1.  

We analyzed our reduced light curves by combining Fourier analysis and least-squares fits in a standard manner using Period04 (see \citet{len05} for program details).  As our two observing runs were separated by a month, the light curves were analyzed on a night-by-night basis.  The amplitude spectra are displayed in Figure 2.  We used a prewhitening technique to perform our temporal analysis, fitting the largest signal in the amplitude spectrum, subtracting the fit from the data, and re-calculating the Fourier transform of the residual light curve.  We repeated this process until the candidate peaks had confidence levels below 99\%, as given by the statistical test proposed by \citet{koe90}.  After each iteration we applied a non-linear, least-squares fitting routine to simultaneously fit the periods, amplitudes, and phases of the isolated frequencies.  Figure 3 shows the amplitude spectra at various stages of this iterative process.  The extracted frequencies are printed in Table 2 and are each labeled \textit{f$_{j}$}, where \textit{j} is ordered in terms of decreasing average amplitude.  The errors shown are derived from the least-squares fits; the actual errors may be larger by a factor of three (see \citealt{mon99}).

The multiperiodic nature of JL 166 is apparent from the amplitude spectrum of the complete light curve.  We report 10 independent oscillations with periods ranging from 97 s to 178 s and amplitudes from 0.9 to 4 mma; the least significant of these has a false alarm probability near 10$^{-6}$.  Some of the reported frequencies were only detected on one of the two nights.  Amplitude variability is detected in many sdB stars observed over an extended period; see for example \citet{ree07}.  Whether the non-detection of some of the modes on one of the two nights is due to the amplitudes decreasing below detection limits is unclear since the amplitudes of the majority of the modes varied between the observation sets.  Most of the power in the spectrum is concentrated between 6800 and 7600 $\mu$Hz.  The highest-frequency variation (13461 $\mu$Hz) is a combination frequency of the dominant peak, \textit{f$_{1}$}, and another lower-amplitude frequency, \textit{f$_{9}$}.  Interestingly, neither \textit{f$_{9}$} nor the cross-frequency is present in the October data.

We note that additional oscillation modes may be present in the amplitude spectrum.  The mean noise level in the 4 to 14 mHz range is greater than that on either side of this band.  This effective plateau in the amplitude spectrum is visible both before and after prewhitening and suggests the presence of additional frequencies that cannot be identified in our data.    

\section{Spectroscopy}

To improve upon the spectroscopy of \citet{heb86}, we obtained four medium-resolution spectra of JL 166 with a combined integration time of 1800 s using the Goodman Spectrograph on 2008 September 19.  We used a 1.03 arcsecond slit and the 600 mm$^{-1}$ VPH grating (0.65 \AA\ pixel$^{-1}$ dispersion) to cover a spectral range from 3500 to 6300 \AA\ with a resolution of 4.5 \AA.  Our spectral images were binned 2x2, resulting in 3.4 pixels per resolution element and a signal-to-noise ratio of approximately 70, measured at the continuum near 4200 \AA.  Table 3 summarizes the details of our spectroscopic observations.

We reduced our spectra with IRAF using standard bias-subtraction and wavelength-calibration procedures.  A standard flat-field correction could not be applied due to a lack of proper flat frames.  We corrected this potential issue by applying a flux calibration using a spectrophotometric standard star (EG 21) whose spectrum was aligned on the same pixels as that of the JL 166 spectra.  The reduced spectral frames were averaged to produce the single, time-averaged spectrum for JL 166 shown in Figure 4.  As expected, our mean spectrum is dominated by Balmer lines, He I lines, and the 4686 \AA\ He II line, confirming the sdOB classification assigned by \citet{heb86}.  We derive a redshift of 79 $\pm$ 27 km s$^{-1}$ from these lines, also in agreement with the results of \citet{heb86}.  

Although typical indications of a low-mass main sequence companion (Ca I triplet, Mg I triplet, \textit{G} band) are absent from the spectrum, Na D absorption lines (5890 \AA\ and 5896 \AA) are present and confirm the existence of a cool companion (see inset of Figure 4).  An over-subtraction of the Na D sky emission features is unlikely as the removal of the OI sky emission line at 5577 \AA\ left no residuals.  Moreover, the Na features in our spectra have redshifts comparable to those computed from the H and He features, and, consequently, we believe they are real.  Contamination from interstellar Na absorption cannot be completely ruled out but is unlikely due to the high galactic latitude (-69.8$^{\circ}$) of JL 166.  As no other spectral features typical of a cool companion were observed, we cannot classify the companion using our data.  The optical-IR color-color plots of \citet{sta04} suggest the companion is an M-type star contributing less than 5\% of the total flux (in the V-band) from the system.   

To determine the atmospheric parameters of JL 166, we employed the model grid of synthetic optical spectra described in \citet{han07}.  The grid is defined in terms of 21 values for the effective temperature from 20000 K to 40000 K (in steps of 1000 K), 11 values for the surface gravity from log \textit{g} of 5.0 to 7.0 dex (in steps of 0.2 dex), and 4 values of the helium-to-hydrogen ratio from log \textit{N}(He)/\textit{N}(H) of -3 to 0 dex (in steps of 1 dex).  Further resolution in the models was achieved through a standard grid interpolation routine \citep{pre07}.  We degraded each model spectrum by convolving it with a Gaussian whose FWHM mimicked that of a single resolution element in our spectrum.  After re-binning the convolved models (while conserving flux) to match the binning of our data, we fit the available absorption lines in our mean spectrum using a least-squares analysis.  

No parameter set consistent with both the H Balmer lines and He ionization equilibrium could be obtained.  This so-called ``Balmer line problem'' is not uncommon amongst sdB stars \citep{heb99}.  The simultaneous fitting of H Balmer lines results in T$_{eff}$ = 33300 K and log \textit{g} = 5.75.  Since the HeI/HeII ratio is highly sensitive to temperature and the He lines are relatively unaffected by gravity, we can also determine T$_{eff}$ by simultaneously fitting the He lines from both ionizations states at the gravity fixed by the Balmer line fits.  This ionization equilbrium indicates T$_{eff}$ = 35400 K and log \textit{N}(He)/\textit{N}(H) = -0.8.  As a compromise, we adopt the average parameters deduced from the He ionization equilibrium and the H Balmer lines:  T$_{eff}$ = 34350 $\pm$ 1000 K, log \textit{g} = 5.75 $\pm$ 0.23, and log \textit{N}(He)/\textit{N}(H) = -0.8 $\pm$ 0.3.  Figure 5 shows the best fits to the H and He profiles.

To illustrate the discrepancies in the model fits, Figure 6 presents the gravity and temperature values resulting from the individual fits to the H Balmer lines, the simultaneous fit to the H Balmer lines, and the He ionization equilbrium requirement.  The H Balmer line fits span a temperature range from 32300 to 33200 K and a log \textit{g} range from 5.4 to 6.3 dex.  A trend of increasing log \textit{g} with increasing excitation level is apparent in the model fits.  The temperature fits, on the other hand, display no obvious trend.  The simultaneous fit to the H$\beta$ to H$\eta$ lines shows that the temperature fit is most regulated by the lower-excitation lines.  The location of the gravity fit is near the average value but is slightly partial to the higher-excitation lines.  As previously noted, the He ionization equilibrium condition requires a temperature around 2000 K hotter than that resulting from the Balmer line fits.  This result is consistent with the findings of \citet{heb99} and has yet to be clearly resolved.

\section{Discussion}

We have discovered a new member of the sdBV pulsator class, JL 166, from high signal-to-noise photometry.  Our spectroscopic observations confirm a sdOB classification and find the spectrum best fits a hot subdwarf model with T$_{eff}$ = 34350 $\pm$ 1000 K, log \textit{g} = 5.75 $\pm$ 0.23 dex, and log \textit{N}(He)/\textit{N}(H) = -0.8 $\pm$ 0.3.  These parameters are in agreement with those cited by \citet{heb86}.  The infrared excess in the J, H, and K bands and our detection of weak Na D lines suggest the presence of a cool companion, probably an M-type star.  Although the contamination from the companion appears to be low (Mg I, Ca I, and the G-band are not visible), we note that our spectroscopic fits are limited by imcomplete knowledge of contamination by the companion.  

The analyses of our photometric observations show that JL 166 is a multimode pulsator, with at least 10 independent frequencies detected between and 5601 $\mu$Hz (178 s) and 10344 $\mu$Hz (97 s) with amplitudes from 0.9 to 4 mma; the most dominant signal is at 134 s.  These oscillations are consistent with p-mode pulsations in hot subdwarf B stars and appear to follow the log \textit{g}-P relation of \citet{koe99}.  Moreover, the frequencies, amplitudes, and number of the observed modes seem comparable to other sdBVs with similar atmospheric parameters.  PG 1047+003, most notably, has T$_{eff}$ and log \textit{g} values comparable to JL 166 and pulsates with periods from 104 to 175 s \citep{kil02}. 

The amplitudes of the detected frequencies differ between observation sets.  These variations are either intrinsic to the star or are due to beating effects from unresolved frequencies. The residuals in the amplitude spectra after prewhitening suggest the presence of additional, unresolved frequencies.  Intrinsic amplitude variations cannot be ruled out either, though, as they are observed in other sdBVs.  Some of the oscillation modes in PG 1047+003, for example, undergo considerable amplitude variations over a timescale of a few weeks \citep{odo98}.  More intensive photometric campaigns are needed to increase the resolution in the amplitude spectrum and distinguish between the two possibilities. 

Cross-correlation of the four spectra obtained showed the radial velocity of the JL 166 sdB component to have changed by 0.0 $\pm$ 5 km s$^{-1}$ in 0.02 days.  No spectroscopic evidence which would justify interpretting the \textit{f$_{1}$} frequency change, between 2008 September 17 and October 14, as a Doppler shift is therefore presently available.  Furthermore, the corresponding change in \textit{f$_{2}$} could not be explained by a Doppler shift which might explain the \textit{f$_{1}$} variation.

\acknowledgements
We acknowledge the support of the National Science Foundation, under award AST-0707381, and are grateful to the Abraham Goodman family for providing the financial support that made the spectrograph possible.  We thank the Delaware Asteroseismic Research Center for providing the filter used in these studies.  We also recognize the observational support provided by the SOAR operators Alberto Pasten, Patricio Ugarte, Sergio Pizarro, and Daniel Maturana.  The SOAR Telescope is operated by the Association of Universities for Research in Astronomy, Inc., under a cooperative agreement between the CNPq, Brazil, the National Observatory for Optical Astronomy (NOAO), the University of North Carolina, and Michigan State University, USA.  The authors are also grateful to an anonymous referee for helpful comments.

\clearpage

\begin{table}
\caption{Time-series Photometry Log}
\begin{tabular}{ccccccc}
\hline
\hline
UT Date & Start Time & T$_{exp}$ & T$_{cycle}$ & Length & Filter\\
(2008) & (UTC) & (s) & (s) & (hr) & \\
\hline
Sep 17 & 00:57:56 & 10 & 13 & 2.21 & S8612\\
Oct 14 & 00:00:00 & 10 & 13 & 2.22  & S8612\\
\hline
\end{tabular}
\label{table:obs}
\end{table}

\clearpage

\begin{table}
\caption{Periodicites detected in the light curve of JL 166.}
\begin{tabular}{ccccc}
\hline
\hline
ID & UT Date & Period & Frequency  & Amplitude \\
   & (2008) &(s) & ($\mu$Hz) & (mma)\\
\hline
\textit{f}1 & Sep 17 & 134.21 $\pm$ 0.06  &  7451.0 $\pm$ 3.1  &  4.08 $\pm$ 0.17 \\
 & Oct 14  &  133.41 $\pm$ 0.09  &  7495.7 $\pm$ 4.9  &  3.33 $\pm$ 0.22  \\
\textit{f}2 & Sep 17 & 146.40 $\pm$ 0.34  &  6831 $\pm$ 16 & 2.90 $\pm$ 1.0 \\
 & Oct 14  &  144.08 $\pm$ 0.31  &  6941 $\pm$ 15  &  1.58 $\pm$ 0.22  \\
\textit{f}3 & Sep 17 & 131.36 $\pm$ 0.16  &  7612.8 $\pm$ 9.5  &  1.35 $\pm$ 0.17\\
 & Oct 14  &  130.66 $\pm$ 0.13  &  7653.5 $\pm$ 7.6  &  2.12 $\pm$ 0.22  \\
\textit{f}4 & Sep 17 & 140.82 $\pm$ 0.14  &  7101.1 $\pm$ 7.0  &  1.85 $\pm$ 0.18 \\
 & Oct 14  &  141.07 $\pm$ 0.29  &  7089 $\pm$ 14  &  1.49 $\pm$ 0.23  \\
\textit{f}5 & Sep 17 & 147.84 $\pm$ 0.52  &  6764 $\pm$ 24  &  1.89 $\pm$ 1.0 \\
 & Oct 14  &  147.30 $\pm$ 0.33  &  6789 $\pm$ 15  &  1.42 $\pm$ 0.22  \\
\textit{f}6 & Sep 17 & - - & - -  & - - \\
 & Oct 14   &  96.68 $\pm$ 0.10  &  10344 $\pm$ 10  &  1.48 $\pm$ 0.21 \\
\textit{f}7 & Sep 17 & 101.42 $\pm$ 0.12  &  9860 $\pm$ 12  &  0.99 $\pm$ 0.17  \\
 & Oct 14  &  101.38 $\pm$ 0.11  &  9864 $\pm$ 10  &  1.46 $\pm$ 0.21  \\
\textit{f}8 & Sep 17 & - - & - -  & - - \\
 & Oct 14  &  178.53 $\pm$ 0.40  &  5601 $\pm$ 13  &  1.20 $\pm$ 0.21  \\
\textit{f}9 & Sep 17 & 167.35 $\pm$ 0.24  &  5975.5 $\pm$ 8.5  &  1.06 $\pm$ 0.17\\
 & Oct 14  &  - -  &  - -  &  - -  \\ 
\textit{f}10 & Sep 17 & 109.52 $\pm$ 0.16  &  9131 $\pm$ 13  &  0.91 $\pm$ 0.17\\
 & Oct 14  &  - -  &  - -  &  - -  \\
\textit{f}1 + \textit{f}9 & Sep 17 & 74.48 $\pm$ 0.05 & 13426.6 $\pm$ 9.0  &  0.86 $\pm$ 0.17 \\
 & Oct 14 & - -  &  - -  &  - -  \\ 
\hline
\end{tabular}
\label{table:freqsol}
\end{table}

\clearpage

\begin{table}
\caption{Spectroscopy Log}
\begin{tabular}{cccccc}
\hline
\hline
Date & T$_{exp}$ & Grating & Slit & Resolution\\
(HJD) & (s) & (1/mm) & (arcsec) & (\AA)\\
\hline
2454728.75552 & 300 & 600 & 1.03 & 4.5\\
2454728.76029 & 500 & 600 & 1.03 & 4.5\\
2454728.76708 & 500 & 600 & 1.03 & 4.5\\
2454728.77381 & 500 & 600 & 1.03 & 4.5\\
\hline
\end{tabular}
\label{table:obs}
\end{table}

\clearpage

\begin{figure}
\epsscale{1}
\plotone{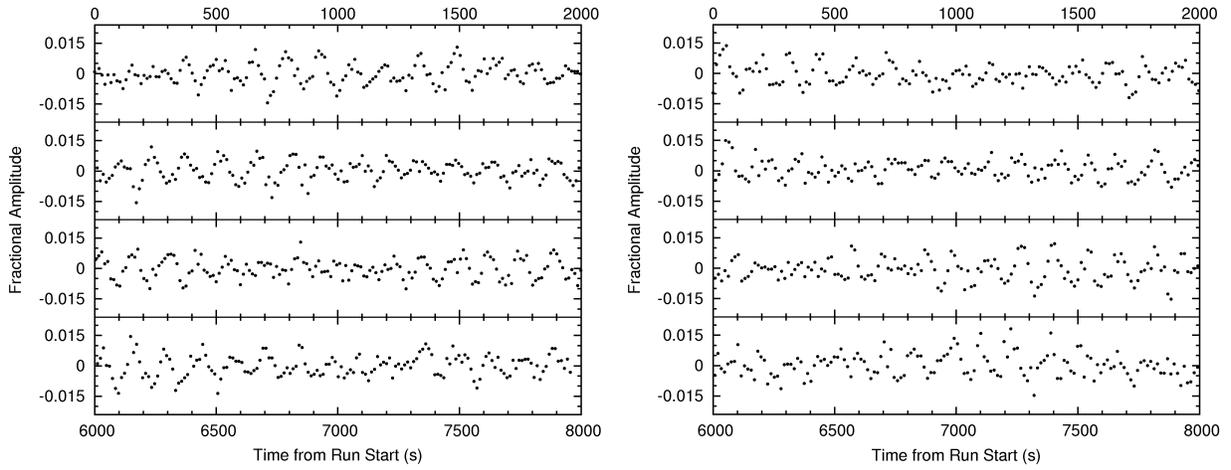}
\caption{Light curves for JL 166 from 2008 September 17 (left panel) and October 14 (right panel).  Small-scale variations in the sky transparency were removed by dividing the light curves by those of constant comparison stars.  The light curves were then normalized with parabolas in order to remove atmospheric extinction effects.  The data are presented unsmoothed.}
\end{figure}

\clearpage

\begin{figure}
\epsscale{0.75}
\plotone{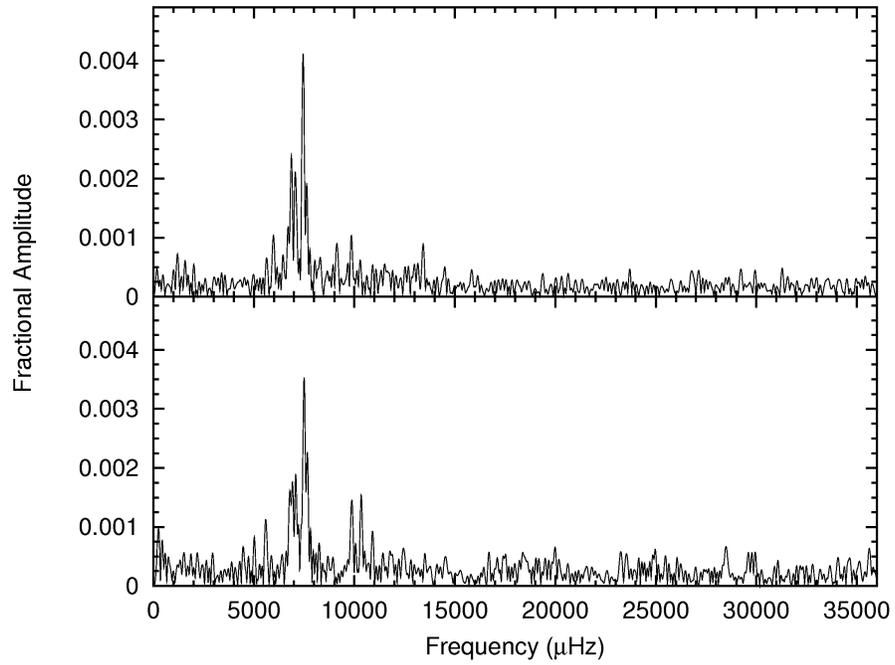}
\caption{Fourier amplitude spectra for the complete light curves from 2008 September 17 (top) and October 14 (bottom) in the 0-35 mHz bandpass.}
\end{figure}

\clearpage

\begin{figure}
\epsscale{1}
\plotone{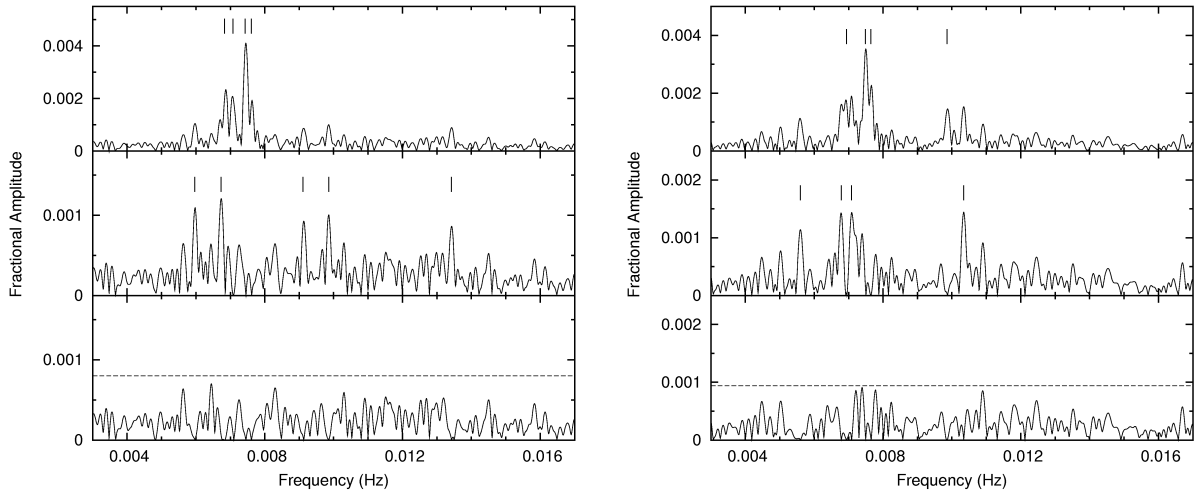}
\caption{Amplitude spectra from 2008 September 17 (left panel) and October 14 (right panel).  Spectra are shown before prewhitening (top), after subtracting the 4 largest peaks (middle), and after subtracting the 9 largest peaks (bottom).  The subtracted peaks in each spectrum are marked in the preceeding amplitude spectra.  Only signals with confidence levels above 99\% (approximately indicated by the dashed lines) are claimed.}
\end{figure}

\clearpage

\begin{figure}
\epsscale{0.75}
\plotone{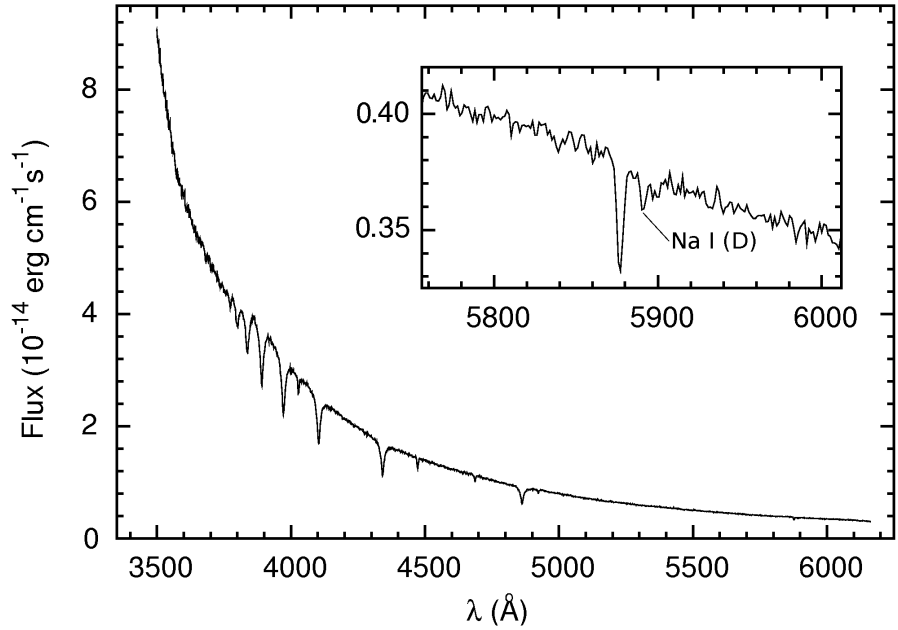}
\caption{Time-averaged, flux-calibrated spectrogram of JL 166.  The spectrum is an average of four 60-s exposures and was flux-calibrated using the spectrophotometric standard star EG 21.  The inset shows a magnification of the reddest portion of the spectrum in which the Na I (D) lines at 5890 \AA\ and 5896 \AA\ are visible next to a He I line as a single, unresolved feature.}
\end{figure}

\clearpage
\begin{figure}
\epsscale{0.75}
\plotone{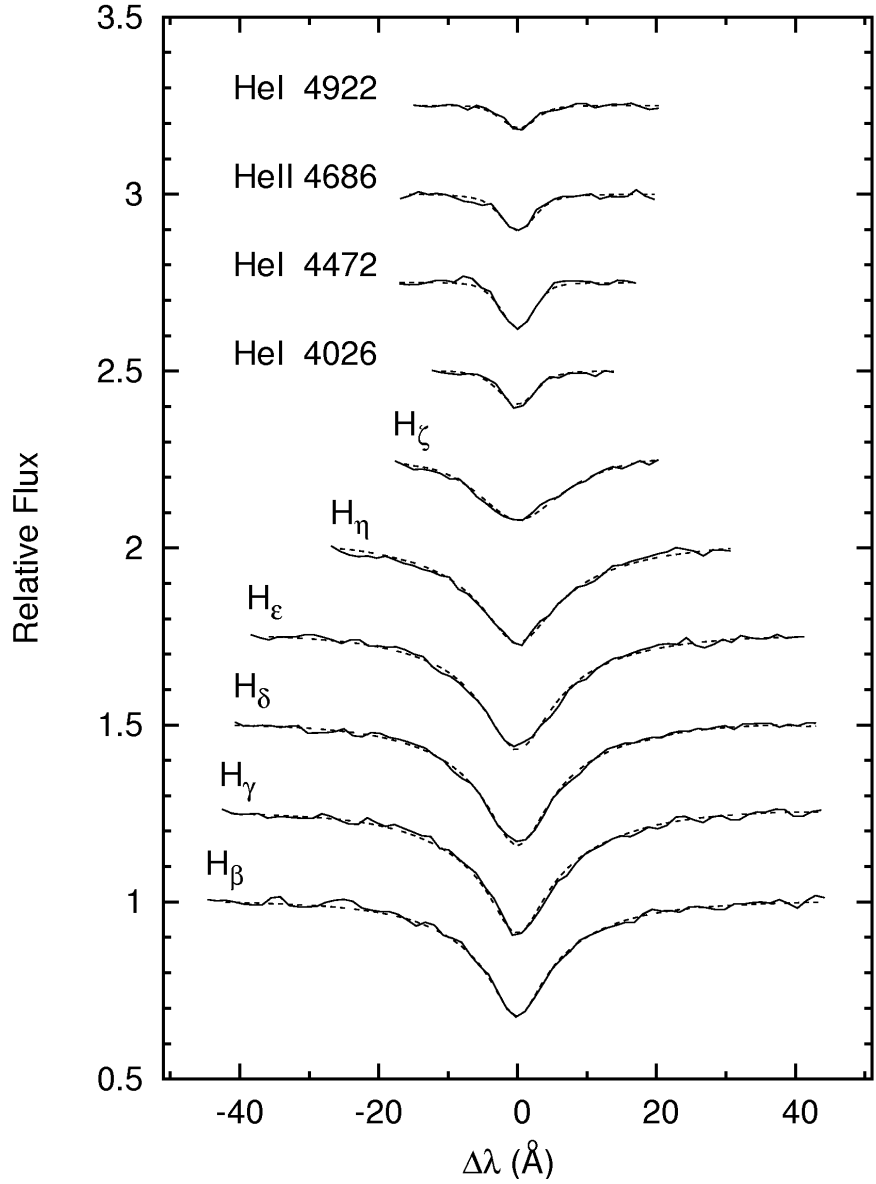}
\caption{Model fits (dotted lines) to the H Balmer and He line profiles in the time-averaged spectrum (solid lines) of JL 166.}
\end{figure}

\clearpage

\begin{figure}
\epsscale{0.75}
\plotone{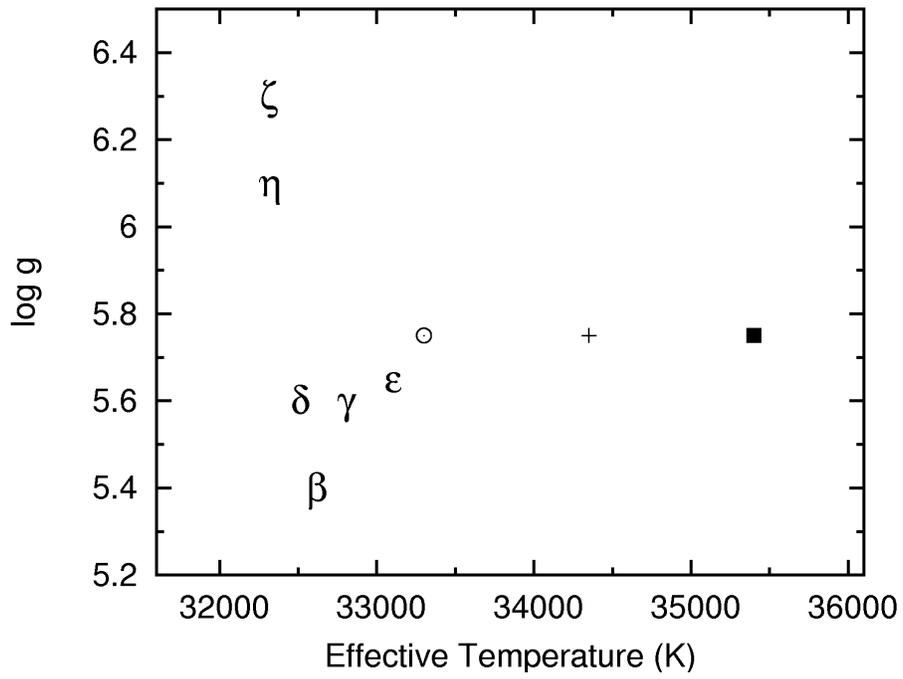}
\caption{Positions of the best individual fits to the H Balmer lines (denoted by their line designations), the best simultaneous fit to the H Balmer lines (open circle), and the He ionization equilibrium solution (filled square) in the log \textit{g}-T$_{eff}$ plane.  Our compromise solution (cross) is also shown and has T$_{eff}$ = 34350 K and log \textit{g} = 5.75.}
\end{figure}

\end{document}